\documentclass{PoS}

\usepackage{amsmath}
\usepackage{amssymb}
\usepackage{MnSymbol}
\usepackage{subfigure}

\title{Phenomenology of the exclusive rare semileptonic decay $\bar{B}\to\bar{K}\pi\ell^+\ell^-$}

\ShortTitle{Phenomenology of $\bar{B}\to\bar{K}\pi\ell^+\ell^-$}

\author{\speaker{Danny van Dyk}\footnote{SI-HEP-2013-04,QFET-2013-03}\\
    Theoretische Physik 1, Naturwissenschaftlich-Technische Fakult\"at,
    Universit\"at Siegen, Walter-Flex-Stra\ss{}e 3, D-57068 Siegen, Germany\\
    E-mail: \email{vandyk@tp1.physik.uni-siegen.de}}

\abstract{I review the phenomenology of exclusive $\bar{B}\to\bar{K}\pi\ell^+\ell^-$
    decays, with focus on the intermediate resonant $\bar{K}^*(892)$ mode.
    The status of angular observables at both ends of the dilepton mass spectrum
    is revisited. Constraints on the $b \to s$ Wilson coefficients $\wc{7,9,10}$ based on 2011
    data from the $B$ factories, CDF and LHCb are presented.
}

\FullConference{14th International Conference on B-Physics at Hadron Machines\\
April 8-12, 2013\\
Bologna, Italy}

\newcommand{\wc}[2][]{\mathcal{C}^\textnormal{#1}_{#2}}
\newcommand{\op}[1]{\mathcal{O}_{#1}}
\newcommand{\dd}{\textnormal{d}}
\newcommand{\para}{\parallel}
\newcommand{\GeV}{\,\textnormal{GeV}}

\newcommand{\order}[1]{O\left(#1\right)}
\newcommand{\refsec}[1]{section \ref{sec:#1}}
\newcommand{\refeq}[1]{eq.\,(\ref{eq:#1})}
\newcommand{\reffig}[1]{fig.\,\ref{fig:#1}}
\newcommand{\thl}{\theta_\ell}
\newcommand{\thK}{\theta_{K^*}}
\DeclareMathOperator{\im}{Im}

\begin{document}

\section{Introduction}
\label{sec:intro}

Since the first observation of the decay $B\to K\ell^+\ell^-$ by the Belle
Collaboration \cite{Abe:2001dh} --- and the subsequent measurements by BaBar \cite{Aubert:2003cm}, CDF \cite{Aaltonen:2008xf},
and most recently LHCb \cite{Aaij:2013iag}, ATLAS \cite{ATLAS} and CMS \cite{CMS:cwa} --- exclusive rare semileptonic
decays are part of the phenomenologist's toolbox and helpful in constraining effects beyond
the Standard Model. However, from a theorist's point of view the exclusive decays
prove to be challenging.\\[-\medskipamount]

The starting point for a theoretical calculation is given by the
framework of an effective Hamiltonian of the form
\begin{equation}
    \mathcal{H}^\textrm{eff} = -\frac{4 G_{\rm F}}{\sqrt{2}} \frac{\alpha_e}{4\pi} V_{tb} V_{ts}^* \sum_i \wc{i}\op{i} + \order{V_{ub} V_{us}^*} + \text{ h.c.}\,,
\end{equation}
for which we follow the conventions of reference \cite{Bobeth:2012vn}.
When considering only factorizable contributions, the set of local operators
can be restricted to the complete basis of semileptonic operators
\begin{equation}
    \op{i} = [\bar{s} \Gamma_i b] [\bar{\ell} \Gamma^\prime_i \ell]\,,\qquad i=9,9',10,10',S,S',P,P',T,T5
\end{equation}
and the radiative operators $\op{7,7'}$. In the Standard Model (SM), one obtains to Next-to-Next-to-Leading Logarithm (NNLL) \cite{Bobeth:1999mk}
\begin{equation}
    \begin{aligned}
        \wc{7}  & \simeq -0.3\,, &
        \wc{9}  & \simeq +4.3\,, &
        \wc{10} & \simeq -4.2\,,
    \end{aligned}
    \label{eq:smpoint}
\end{equation}
and the Wilson coefficients of the remaining radiative and semileptonic operators
are either suppressed by $m_s/m_b$, $m_\ell/M_W$ or vanish.
The understanding of long-distance effects introduced by four-quark operators $\op{1,\dots,6}$ and the chromomagnetic operator $\op{8}$ are crucial to precision studies of exclusive semileptonic rare $\bar{B}$ decays.\\[-\medskipamount]

In the following I will revisit the calculational approaches
to the decay $\bar{B}\to\bar{K}\pi\ell^+\ell^-$ with special regard to the long-distance contributions
in two regions of the dilepton mass, and discuss the powerful constraints on the
$b\to s$ Wilson coefficients $\wc{7,9,10}$ that can be obtained from experimental data on exclusive radiative
and (semi)leptonic decays.

\section{The Angular Distribution of $\bar{B}\to\bar{K}\pi \ell^+\ell^-$ Decays}

The decay
\begin{equation}
    \bar{B}(p) \to \bar{K}(k_1) \pi(k_2) \ell^+(q_1) \ell^-(q_2),\qquad q = q_1 + q_2, k = k_1 + k_2
\end{equation}
can be fully described by means of five kinematic variables: $q^2$ the dilepton mass squared, $k^2$ the $\bar{K}\pi$ mass squared,
the angle $\theta_\ell$ between the $\ell^-$ momentum and the $\bar{B}$ momentum in the dilepton rest frame,
the angle $\theta_{K^*}$ between the $\bar{K}$ momentum and the opposite $\bar{B}$ momentum in the $\bar{K}\pi$ rest frame, and 
the angle $\phi$ between the $\bar{K}\pi$ and dilepton decay planes. The differential decay width can then
be decomposed into 18 angular observables $J_n \equiv J_n(q^2, k^2)$ \cite{Bobeth:2012vn,Kruger:2005ep,Becirevic:2012dp,Blake:2012mb}
\begin{equation}
\begin{split}
  \frac{8 \pi}{3} & \frac{d^4 \Gamma}{d q^2\, d k^2\, d\!\cos\thl\, d\!\cos\thK\, d\phi} = \\
      & (J_{1s} + J_{2s} \cos\!2\thl + J_{6s} \cos\thl) \sin^2\!\thK
    + (J_{1c} + J_{2c} \cos\!2\thl + J_{6c} \cos\thl) \cos^2\!\thK\\\
    + & (J_{1i} + J_{2i} \cos\!2\thl) \cos\!\thK
    + (J_3 \cos 2\phi + J_9 \sin 2\phi) \sin^2\!\thK \sin^2\!\thl\\
    + & (J_4 \cos\phi + J_8  \sin\phi) \sin 2\thK \sin 2\thl 
    + (J_{4i} \cos\phi + J_{8i}  \sin\phi) \sin \thK \sin 2\thl\\
    + & (J_5 \cos\phi  + J_7 \sin\phi ) \sin 2\thK \sin\thl
    + (J_{5i} \cos\phi  + J_{7i} \sin\phi ) \sin \thK \sin\thl \,.
\end{split}
    \label{eq:angdist}
\end{equation}
The distribution \refeq{angdist} is sufficient to describe model-independently
the effects of the complete basis of $b\to s\ell^+\ell^-$ operators as given in \refsec{intro}. It also incorporates
pure P-wave states ($n=3,4,5,6s,6c,7,8,9)$, combined P- and S-wave states ($n=1s,1c,2s,2c$), as well as S-P interference terms ($n=1i,2i,4i,5i,7i,8i$) \cite{Becirevic:2012dp,Blake:2012mb}.\\[-\medskipamount]

Since the hadronic matrix elements for non-resonant S-wave $B\to K \pi$ transitions are not yet sufficiently understood, studies of the their interference effects rely on calculations
based on resonance models \cite{Becirevic:2012dp}, or resort to extraction of the S-wave contributions from data \cite{Blake:2012mb}.
So far most calculations, however, assume a resonant on-shell P-wave $K^*(892)$ state only, which is subsequently handled in a narrow width
approximation \cite{Kruger:2005ep,Altmannshofer:2008dz}. The latter restricts $k^2 = M_{K^*}^2$, thereby reducing the
number of independent kinematic variables from five to four. The hadronic matrix elements for such $\bar{B}\to\bar{K}^*$ transitions
are traditionally expressed in terms of seven $q^2$-dependent form factors $V,A_{0,1,2},T_{1,2,3}$. In the heavy quark limit one obtains the Isgur-Wise relations \cite{Isgur:1990kf}
between the dipole and vector form factors. Using the traditional form factor convention these relations read
\begin{gather}
    \nonumber
    \begin{aligned}
    T_1(q^2) & = \kappa(\mu) V(q^2)\,, &
    T_2(q^2) & = \kappa(\mu) A_1(q^2)\,,
    \end{aligned}\\
    T_3(q^2) = \frac{\kappa(\mu)M_B}{q^2} \big((M_B - M_{K^*}) A_2(q^2) - (M_B + M_{K^*}) A_1(q^2)\big)
    \label{eq:iwr}
\end{gather}
up to corrections $\order{\Lambda_\textnormal{QCD}/m_b}$ with $\kappa(m_b) = 1 + \order{\alpha_s^2}$ \cite{Bobeth:2010wg}.
They reduce the number of independent form factors to four \emph{helicity form factors} \cite{Bobeth:2010wg,Bharucha:2010im}
\begin{equation}
    \begin{aligned}
        f_S     & = \frac{2\sqrt{\lambda}}{\sqrt{q^2}} A_0 &
        f_0     & = \frac{(M_B^2 - M_{K^*}^2 - q^2)(M_B + M_{K^*})^2 A_1 - \lambda A_2}{2 M_{K^*} (M_B + M_{K^*}) \sqrt{q^2}}\\
        f_\perp & = \frac{\sqrt{2\lambda}}{M_B + M_{K^*}} V\qquad &
        f_\para & = \sqrt{2} (M_B + M_{K^*}) A_1\,.
    \end{aligned}
    \label{eq:helff}
\end{equation}
with the K\"all\'en function $\lambda = \lambda(M_B^2, M_{K^*}^2, q^2)$.
Additionally, in the Large Energy Limit (LEL) for $E_{K^*}$, the $\bar{K}^*$ energy
in the $\bar{B}$ rest frame, the four helicity form factors reduce to two universal soft form factors:
$\xi_\perp$ and $\xi_\para$ \cite{Charles:1998dr,Beneke:2000wa}.\\[-\medskipamount]

The form factors are, nevertheless, inherently non-perturbative
quantities, and as such can only be calculated through non-perturbative methods.
For $q^2 \simeq 0$, Light Cone Sum Rules (LCRSs) provide access \cite{Ball:2004rg,Khodjamirian:2010vf}.
On the other hand, Lattice QCD (LQCD) can -- in principle -- be used to obtain the form factors numerically at
large $q^2 \gtrsim 15\GeV^2$ \cite{Liu:2011raa}. Application of a series expansion based on the analytic
properties of the form factors in the complex plane allows to extrapolate the LCSR and LQCD results to
intermediate $q^2$ values \cite{Bharucha:2010im}. This approach can be further improved by using
experimental data to constrain ratios $f_\perp/f_\para$, $f_0/f_\para$ \cite{Hambrock:2012dg,Beaujean:2012uj,Hambrock:2013xxx}.\\[-\medskipamount]

If one considers only contributions of semileptonic operators $[\bar{s}\Gamma b][\bar\ell \Gamma^\prime \ell]$, the angular
observables $J_n$ take a simple form and can be expressed in terms of 14 complex-valued transversity amplitudes $A_{k}$ \cite{Bobeth:2012vn},
cf.\ eqs. (B1)-(B12) of reference \cite{Bobeth:2012vn}. This ansatz of \emph{naive factorization}
is known to be broken by the peaking background of processes $\bar{B}\to\bar{K}\pi \psi(1S,2S,\dots)(\to \ell^+\ell^-)$. The narrow charmonium resonances $\psi(1S)$ and $\psi(2S)$ are cut from the experimental data by means of
two $q^2$ vetoes. However, their tails as well as broader charmonium resonances still affect the theoretical predictions. It is therefore illstrustrative to compute the results in the naive factorization ansatz, and systematically extend it with known corrections.\\[-\medskipamount]

In order to incorporate non-factorizing effects into the description of the
$\bar{B}\to\bar{K}\pi\ell^+\ell^-$ amplitudes, one turns to study the correlation function
\begin{equation}
    \mathcal{T}^\mu_i = i \int \dd x\, e^{i q x}\,\langle \bar{K}^* | T\lbrace \op{i}(0) j^\mu_\textnormal{e.m.}(x) |\bar{B}\rangle
    \label{eq:correlator}
\end{equation}
for the four-quark and chromomagnetic operators $\op{1,\dots,6,8}$. Here $T$ is the time-ordered product and $j_\textnormal{e.m.}$
is the electromagnatic current. Since the quantum numbers of the correlator \refeq{correlator} are compatible
only with the operators $\op{7('),9(')}$, one can account for intermediate $\bar{q}q$ effects by means of
the replacements
\begin{align}
    \wc{i} \to \wc{i} + \Delta_i(q^2) \equiv \wc[eff]{i}(q^2)\,,\qquad i=7('),9(')
\end{align}
where $\wc[eff]{i}(q^2)$ are known as the $q^2$-dependent \emph{effective Wilson coefficient}. Note here
that the $\Delta_i(q^2)$ in general include both factorizing, form factor independent as well as non-factorizing,
form factor dependent contributions. In the following
paragraphs two major approaches to the calculation of \refeq{correlator} shall be revisited.\\[-\medskipamount]

For small $q^2$, or equivalently for large recoil energy $E_{K^*} \sim m_b$ of the $\bar{K}^*$
in the $\bar{B}$ rest frame, one can systematically calculate certain perturbative contributions to the
effective Wilson coefficients in the framework of QCD Factorization (QCDF) \cite{Beneke:2001at,Beneke:2004dp}.
Schematically, one obtains for projections of the correlator $\mathcal{T}_a$
\begin{equation}
    \mathcal{T}_a \supseteq C_a \xi_a + \phi_B \otimes T_a \otimes \phi_{K^*} + \order{\frac{\Lambda_\textnormal{QCD}}{m_b}} \,,\qquad a=\perp,\para\,,
\end{equation}
where the $C_a$ denotes factorizable corrections that enter with the soft form factors $\xi_a$. The
hard scattering kernels $T_a$ enter after convolution with the $\bar{B}$ and $\bar{K}^*$ light cone
distribution amplitudes (LCDAs) $\phi_B$ and $\phi_{K^*}$, respectively. The calculation of some power-suppressed
contributions is plagued by endpoint divergencies \cite{Beneke:2004dp}, a problem that can potentially
be overcome with the help of LCSR methods \cite{Dimou:2012un,Lyon:2013gba}. LCSR methods also allow access to further corrections beyond
QCDF \cite{Khodjamirian:2010vf,Jager:2012uw}.\\[-\medskipamount]

On the other end of the $q^2$ spectrum, where $q^2 \simeq m_b^2$, a local Operator Product Expansion (OPE) can be
performed simultaneosuly in $1/m_b$ and $1/\sqrt{q^2}$ \cite{Grinstein:2004vb,Beylich:2011aq}.
Combining the results of the OPE with the form factor relations \refeq{iwr}, one finds \cite{Bobeth:2010wg} that the correlator \refeq{correlator}
yields only factorizable contributions to the transversity amplitudes up to corrections $\order{\alpha_s \frac{\Lambda}{m_b}, \frac{\wc{7}}{\wc{9}}\frac{\Lambda}{m_b}}$, where $\Lambda = \order{\Lambda_\textnormal{QCD}}$.
Thus, the transversity amplitudes $A_{0,\perp,\para}^{L(R)}$ factorize to that order,
\begin{align}
    A_{0,\para}^{L(R)}(q^2) & = -\wc{-,L(R)}(q^2) \times f_{0,\para}(q^2)\,, &
    A_{\perp}^{L(R)}(q^2)   & = +\wc{+,L(R)}(q^2) \times f_\perp(q^2)\,,
\end{align}
into helicity form factors $f_{0,\perp,\para}$ and effective short-distance coefficients $\wc{\pm,L(R)}$.
(In the SM basis, i.e., for $\wc{9',10'} = 0$ one obtains $\wc{+,L(R)} = \wc{-,L(R)}$).
Matrix elements other than those parametrized by the $\bar{B}\to\bar{K}^*$ form factor enter
only at order $\Lambda^2 / m_b^2$ \cite{Beylich:2011aq}. Subleading corrections to the transversity
ampltiudes $A_k$ are parametrically suppressed and can be parametrized as \cite{Bobeth:2011gi}
\begin{equation}
    r_{k} = \frac{\Lambda_k}{M_B}\big(\wc{7} + \alpha_s e^{i\delta_k}\big)\,.
\end{equation}
Beyond probing electro-weak and hadronic quantities,
observables at low hadronic recoil also provide quantitative tests of the OPE. The latter
are constructed from terms $\varepsilon_k$ that could break the factorization of the transversity
amplitudes $A_k$,
\begin{equation}
    A^{L(R)}_k \propto \wc{L(R)} f_k (1 + \varepsilon_k)\,,\qquad k=0,\perp,\para\,.
\end{equation}
Based on this ansatz, one finds that $J_7 \neq 0$ probes $\im{\varepsilon_{0,\para}}$ (and thus the OPE)
at the percent level, while contributions from Beyond the SM (BSM) are helicity suppressed \cite{Bobeth:2012vn}.\\[-\medskipamount]

\begin{figure}
    \hspace{.05\textwidth}
    \subfigure[Performance of $H_T^{(2,3)}$ vs $A_{\rm FB}$ in probing the ratio $|\wc{9}/\wc{10}|$. Both observables are shown including their theory uncertainty for the entire low recoil bin, and normalized to their respective
        SM values. \label{fig:np-sens1}]{\includegraphics[width=.38\textwidth]{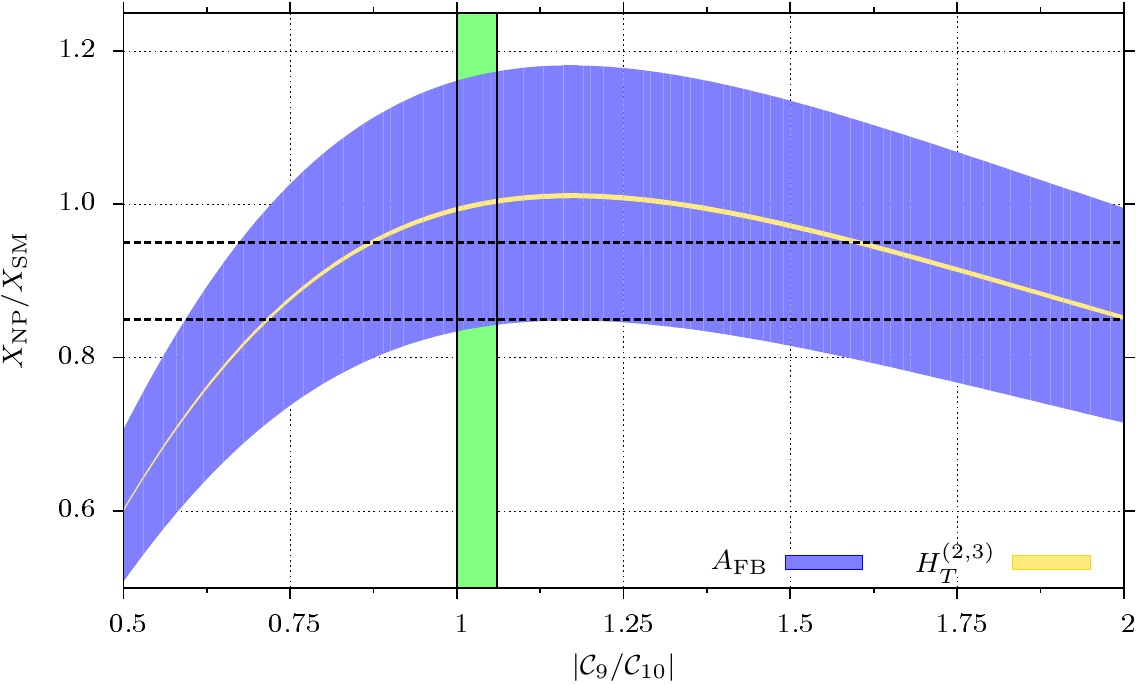}}\hfill
        \subfigure[Performance of varios CP asymmetries as functions of $\im{\wc{10'}}$. Of the compared
        observables, $a_{CP}^{(4)}$, the CP asymmetry of $H_T^{(4,5)}$, is most sensitive to small deviations
    of the SM value $\im{\wc{10'}} = 0$. \label{fig:np-sens2}]{\includegraphics[width=.38\textwidth]{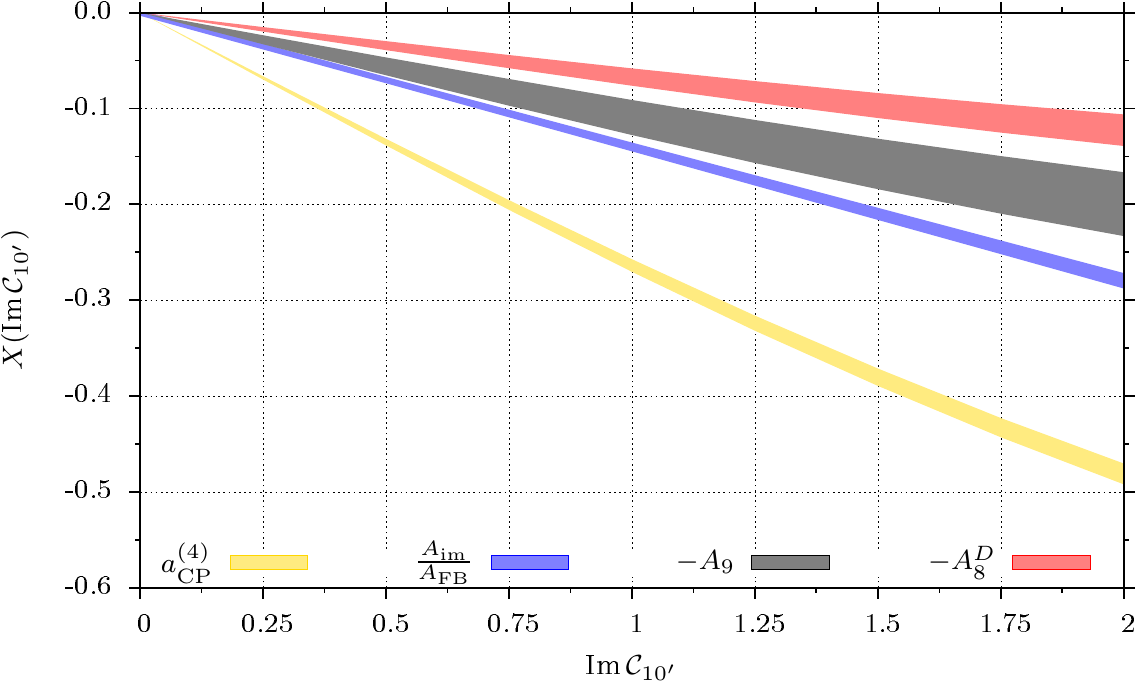}}\hspace{.05\textwidth}\,\!
\caption{Both figures taken from reference \cite{Bobeth:2012vn}.}
\end{figure}
The theoretical uncertainties in $\bar{B}\to(\bar{K}\pi)_P\ell^+\ell^-$ observables
are driven by the lack of knowledge of the $\bar{B}\to\bar{K}^*$ form factors. This
dilutes the constraining power of experimental data on observables such
as the branching ratio $\mathcal{B}$, the longitudinal polarization $F_L$ and the
lepton forward-backward asymmetry $A_{\rm FB}$. In order
to reduce the impact of form factor uncertaintes, several analyses \cite{Bobeth:2012vn,Bobeth:2010wg,Egede:2008uy,Egede:2010zc,Becirevic:2011bp} have been
carried out that introduce \emph{optimized observables}. The latter are designed so that
form factors cancel to leading order.\\
For the region of low hadronic recoil, five observables $H_T^{(i)}$,
\begin{equation}
    \begin{aligned}
        H_T^{(1)} & = \frac{\sqrt{2} J_4}{\sqrt{-J_{2c}(2J_{2s} - J_3)}}\,, &
        H_T^{(2)} & = \frac{\beta_\ell J_5}{\sqrt{-2J_{2c}(2J_{2s} + J_3)}}\,, &
        H_T^{(3)} & = \frac{\beta_\ell J_{6s}}{2\sqrt{4J_{2s}^2 - J_3^2}}\,,\\
        H_T^{(4)} & = \frac{2 J_8}{\sqrt{-2J_{2c}(2J_{2s} + J_3)}}\,, &
        H_T^{(5)} & = \frac{-J_{9}}{\sqrt{4J_{2s}^2 - J_3^2}}\,,\\
    \end{aligned}
\end{equation}
can be constructed that fulfill the cancellation requirement. While $|H_T^{(1)}| \simeq 1$
almost model indepedently, $H_T^{(2,3)}$ are effective in probing the ratio $|\wc{9}/\wc{10}|$ \cite{Bobeth:2012vn};
their increased BSM sensitivity in comparison to $A_{\rm FB}$ can be inferred from \reffig{np-sens1}.
Finally, $H_T^{(4,5)}$ and their CP asymmetries probe CP-violating right-handed Wilson coefficients
$\wc{9',10'}$, see \reffig{np-sens2}. As a consequence of the LEL all of the $H_T^{(i)}$ stay also free of form factors to leading order at large hadronic recoil \cite{Descotes-Genon:2013vna}. A complete basis of optimized observables
for the large recoil region, with focus on the extraction of $\wc{7'}$ is also given in reference \cite{Descotes-Genon:2013vna}.

\section{Constraining $b\to s\gamma$ and $b\to s\ell^+\ell^-$ Wilson Coefficients from Exclusive Decays}

\begin{figure}
\hspace{.1\textwidth}
\includegraphics[width=.38\textwidth]{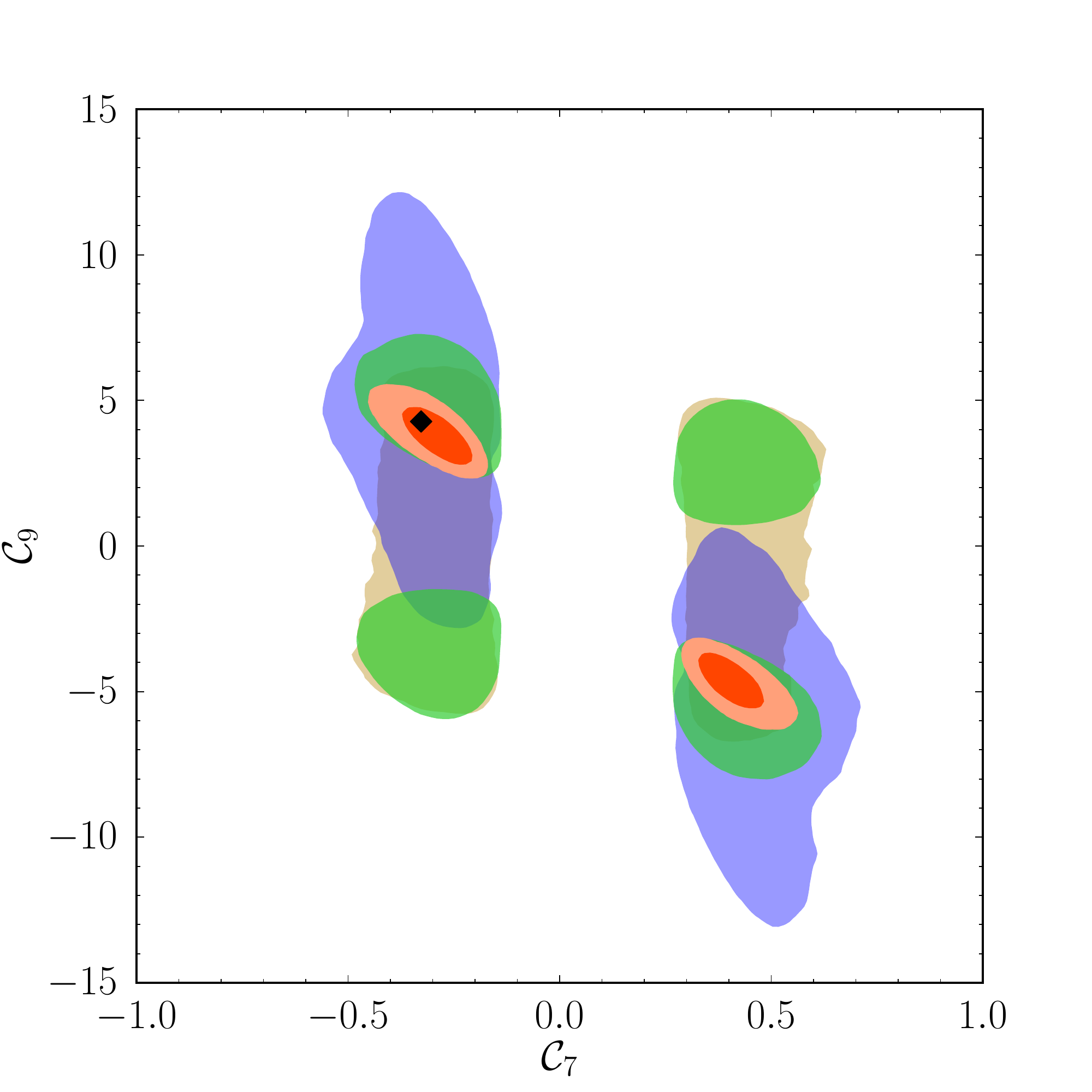}\hfill
\includegraphics[width=.38\textwidth]{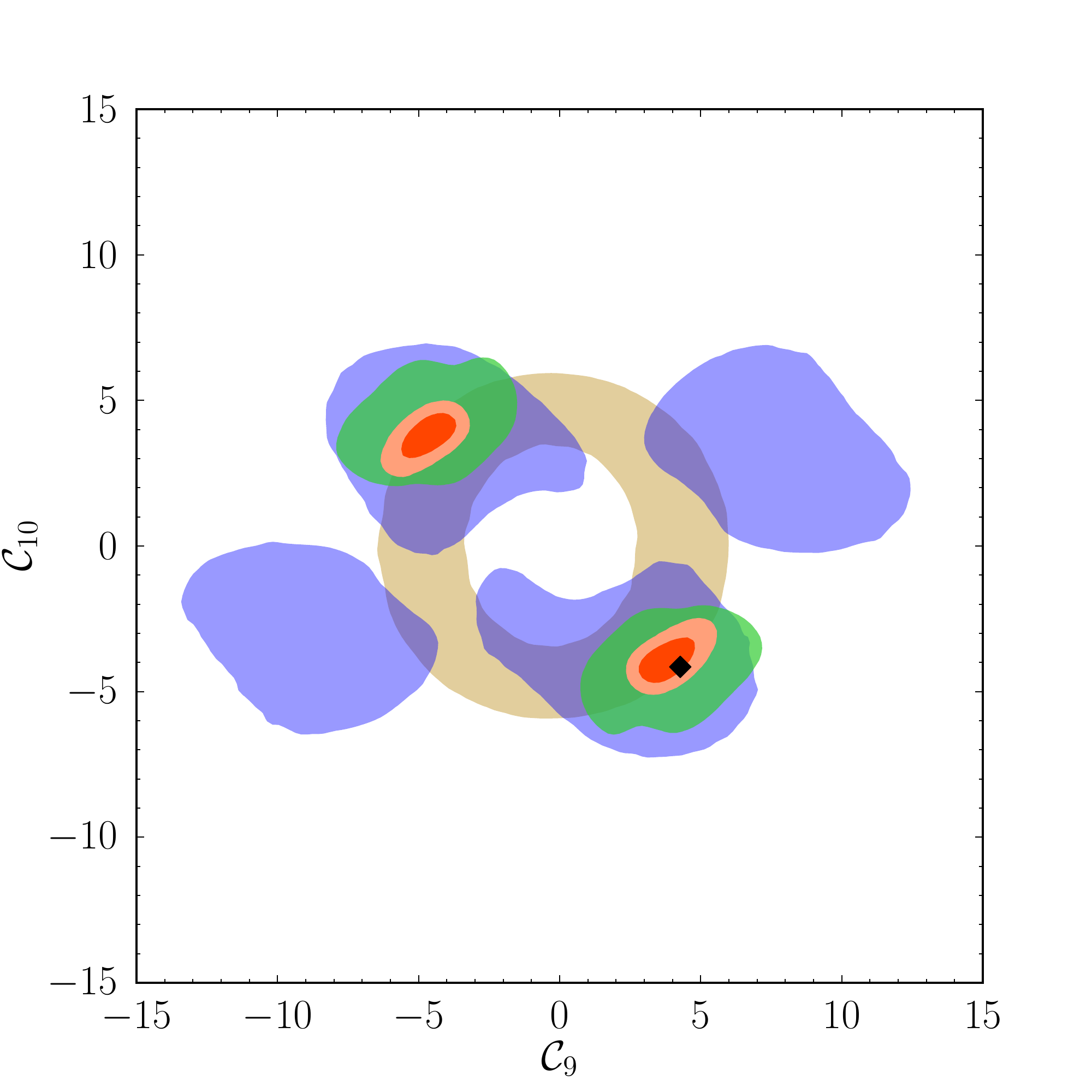}
\hspace{.1\textwidth}\,\!
\caption{SM values ($\filleddiamond$) and $95\%$ credibility regions for the Wilson coefficients $\wc{7,9,10}$ as obtained from the analysis in reference \cite{Beaujean:2012uj}. \label{fig:overlay}}
\end{figure}

The Wilson coefficients $\wc{i}$ can be constrained in a model-independent
framework, i.e., one fits their values from all available experimental $|\Delta B|=|\Delta S|=1$ data
in a global analysis. For the complete basis of semileptonic operators with --- in general --- complex-valued
Wilson coefficients, this means to perform a fit with 20 real-valued parameters of interest.
Taking into account the Wilson coefficients of four-quark and radiative operators
$\op{1,\dots,6}$ and $\op{7('),8(')}$ respectively on arrives at 40 degrees of freedom.\\[-\medskipamount]

It it therefore necessary to lower the number of parameters of interest. One option
is to restrict the Wilson coefficients $\wc{1,\dots,6,8(')}$ to their SM values, and set
the non-SM-like coefficient $\wc{7',9',10',S,P,T,T5}$ to zero. This scenario corresponds to
the \emph{SM basis of operators} and was considered in several analyses \cite{Beaujean:2012uj,DescotesGenon:2011yn,Altmannshofer:2011gn,DescotesGenon:2012zf}. It was shown in reference \cite{Beaujean:2012uj} that this basis of operators suffices
to describe the available data on exclusive rare (semi)leptonic and radiative decays. I refer the reader
to the original work for details on experimental inputs and the statistical method.
Two best-fit points $(\wc{7},\wc{9},\wc{10})$ are obtained,
\begin{align}
    \text{SM-like:} & (-0.293,+3.69,-4.19)\,, &
    \text{sign-flipped:} & (+0.416,-4.59,+4.05)\,.
\end{align}
Both of these points yield $p$-values of $60\%$ and $75\%$, depending on the definition
of the $p$-value. The SM-like solution is compatible with the SM
point \refeq{smpoint} at $68\%$ credibility
\cite{Beaujean:2012uj}. The $95\%$ credibility regions for $\wc{7,9,10}$ obtained from this analysis
are shown in \reffig{overlay}.\\[-\medskipamount]

Beyond obtaining information on the nature of the Wilson coefficients, the analysis at hand
also provides information on the hadronic quantities such as form factor values. Their treatment
as nuisance parameters in a Bayesian analysis allows to compare their prior and
posterior probability distributions. In the particular analysis, it was informative
to see that a slight ($\lesssim 1\sigma$) tension arises between the priors and posteriors of the
form factors $V_1$ and $A_2$ at $q^2 = 0$ \cite{Beaujean:2012uj}. This behaviour has also been
seen in an analysis dedicated to extracting the form factors from low recoil data \cite{Hambrock:2012dg}.

\section{Conclusion}

The theoretical understanding and the experimental handling of the decay $\bar{B}\to\bar{K}\pi\ell^+\ell^-$ have
both advanced tremendously over the last decade. Still, there is room for improvements. Global analyses
of exclusive rare semileptonic decays find good agreement with the SM, but leave plenty of room for
physics beyonds the SM. The emergence of precision data on exclusive rare semileptonic decays from LHCb
and the prospect of inclusive measurements from Belle II will not only allow to constrain BSM effects further.
It will also aide in improving our understanding of hadronic models and non-perturbative methods.

\section*{Acknowledgments}
I wish to thank the organizers of Beauty 2013 for their invitation, and the opportunity to
present this talk. I am also grateful to Christoph Bobeth and Thorsten Feldmann for comments on the manuscript.
My work presented here is supported in parts by the Deutsche Forschungsgemeinschaft (DFG) within
Research Unit FOR 1873 (``QFET'').

\end{document}